\documentclass[%
aps,
 amsmath,amssymb,
reprint,%
superscriptaddress,floatfix]{revtex4-1}

\usepackage{graphicx}
\usepackage{dcolumn}
\usepackage{bm}
\usepackage{placeins}
\usepackage{color}

\begin{document}

\title{Particle-in-Cell Simulations of Collisionless Magnetic Reconnection with a Non-Uniform Guide Field}

\author{F. Wilson}
\email{Electronic Mail: fw237@st-andrews.ac.uk}
\author{T. Neukirch}
\email{Electronic Mail: tn3@st-andrews.ac.uk}
\affiliation{School of Mathematics and Statistics, University of St Andrews, St Andrews, UK, KY16 9SS}
\author{M. Hesse}
\affiliation{NASA Goddard Space Flight Center, Greenbelt, MD 20771, USA}
\author{M. G. Harrison}
\affiliation{School of Mathematics and Statistics, University of St Andrews, St Andrews, UK, KY16 9SS}
\author{C. R. Stark}
\affiliation{School of Mathematics and Statistics, University of St Andrews, St Andrews, UK, KY16 9SS}
\affiliation{Division of Computing and Mathematics, Abertay University, Dundee, DD1 1HG} 

\begin{abstract}
Results are presented of a first study of collisionless magnetic reconnection starting from a recently found exact nonlinear force-free Vlasov-Maxwell equilibrium.
The initial state has a Harris sheet magnetic field profile in one direction and a non-uniform guide field in a second direction, resulting in a spatially constant magnetic field strength
as well as a constant initial plasma density and plasma pressure.
It is found that the reconnection process initially resembles guide field reconnection, but that a gradual transition to anti-parallel reconnection happens as the system evolves.
The time evolution of a number of plasma parameters is investigated, and the results are compared with simulations starting from a Harris sheet equilibrium and a Harris sheet plus constant guide field equilibrium.
\end{abstract}

\maketitle

\section{Introduction}

Magnetic reconnection is one of the most fundamental plasma processes, and plays an important role in the magnetic activity of many astrophysical and laboratory plasmas \cite{Biskamp-2000,Birn-2007}. 
It allows the conversion of stored magnetic energy into bulk flow, thermal and non-thermal energy, through changes in magnetic connectivity. In many astrophysical plasmas, the effects of particle collisions are negligible, and various aspects of collisionless reconnection have previously been studied in great detail \cite{Kuznetsova-1998,Shay-1998,Hesse-1999,Kuznetsova-2000,Kuznetsova-2001,Hesse-2001b,Pritchett-2001,Hesse-2002,Rogers-2003,Hesse-2004,Ricci-2004,Pritchett-2004,Hesse-2005,Pritchett-2005,Hesse-2006,Daughton-2007,Wan-2008,Daughton-2011,Hesse-2011,Eastwood-2013,Hesse-2013,Aunai-2013b}. One particular aspect which has been investigated by a number of authors (e.g. Refs.~\onlinecite{Pritchett-2001,Hesse-2002,Rogers-2003,Hesse-2004,Ricci-2004,Pritchett-2004,Hesse-2005,Pritchett-2005,Hesse-2006,Wan-2008,Hesse-2011}) is the influence of a guide field on the reconnection process. Most of these studies have used a Harris sheet \cite{Harris-1962} with a constant guide field as an initial current sheet configuration. 


The addition of a constant guide field to the Harris sheet affects the evolution in a number of ways (see e.g. Ref.~\onlinecite{Birn-2007} for a more comprehensive overview than what we describe here). Some important points to note are as follows:
\begin{enumerate}
\item[(a)] A constant guide field (of sufficient magnitude) has been shown to reduce the reconnection rate \cite{Pritchett-2001,Ricci-2004,Hesse-2004}.
\item[(b)] 
The structure of the diffusion region is changed with the addition of a constant guide field \cite{Birn-2007}. In the anti-parallel (Harris sheet) case, the different outflow trajectories of the ions and electrons generate in-plane current loops (Hall currents), which in turn generate a quadrupolar out-of-plane magnetic field \cite{Shay-1998,Hesse-2001b}. The addition of a constant (out-of-plane) guide field results in a distortion of this quadrupolar field \cite{Pritchett-2001,Hesse-2002}. Furthermore, there is a strong parallel component to the out-of-plane electric field, which generates strong out-of-plane currents, and in-plane components of the parallel electron flows produce a density asymmetry along the separatrices.
\item[(c)] As a result of the density asymmetry described in point (b), in guide field reconnection there is a rotation of the reconnecting current sheet(s) \cite{Hesse-2002,Hesse-2004,Pritchett-2004,Hesse-2005,Hesse-2006,Hesse-2011}.
\item[(d)] A guide field affects the particle orbits in the electron diffusion region \cite{Hesse-2011} - it can destroy the bounce motion which occurs across the field reversal in the anti-parallel case, and so the length scales characterising the orbits in each case are different. In the guide field case, the relevant scale is the electron Larmor radius in the guide field, whereas in the anti-parallel case it is the electron bounce width in the reconnecting field component. 
\item[(e)] A consequence of point (d) is that the addition of a guide field leads to thinner current sheets than in the anti-parallel case \cite{Hesse-2006}.
\end{enumerate}





In this paper, we wish to address the following question: does the reconnection process change (and, if so, how?) if we use an initial current sheet configuration with a non-uniform guide field? We present results of a 2.5D particle-in-cell (PIC) simulation, in which we use an exact self-consistent equilibrium for the force-free Harris sheet as an initial condition \cite{Harrison-2009b, Neukirch-2009}. 

Since the equilibrium guide field of the force-free Harris sheet (here $B_y=B_0/\cosh(z/L)$) decreases with distance from the centre of the current sheet, we expect that the system will initially show 
features of guide field reconnection, but that a gradual transition to anti-parallel reconnection should take place, because plasma with smaller guide field strength
should be transported towards the reconnection region as the system evolves in time. 
We will investigate whether and how this transition takes place, and also how it is reflected in the time evolution
of plasma quantities relevant for collisionless reconnection, such as the off-diagonal,
 non-gyrotropic elements of the electron pressure tensor.

Three-dimensional PIC simulations have previously been carried out for a magnetic field profile similar to that of the force-free Harris sheet \cite{Hesse-2005}, but with an additional constant guide field added in the same direction as the non-uniform guide field. The initial particle distribution functions were taken to be drifting Maxwellian distributions, which do not represent an exact initial equilibrium for this configuration. This leads us to discuss another motivation for our work - we are not aware of any previous study of collisionless reconnection for which exactly force-free initial conditions have been used for a nonlinear force-free field. The only known studies to use exactly force-free initial conditions have started from a linear force-free configuration \cite{Bobrova-2001, Li-2003, Nishimura-2003,Sakai-2004,Bowers-2007}. Exact collisionless equilibria for such 1D linear force-free fields were first found approximately five decades ago \cite{Moratz-1966, Sestero-1967}, but the first exact equilibria of this type for nonlinear force-free fields were found only very recently \cite{Harrison-2009b, Neukirch-2009, Wilson-2011, Stark-2012, Abraham-Shrauner-2013,Allanson-2015b}. Hence, only preliminary investigations have been carried out into the linear and nonlinear collisionless stability and dynamics of these configurations \cite{mike-thesis,fionawilson-thesis}.

The structure of the paper is as follows. In Sect. \ref{sec:setup} we discuss the simulation setup, followed by a detailed description of the results in Sect. \ref{sec:results}. We conclude with a summary and conclusions in Sect. \ref{sec:summary}.

\section{Simulation Setup}
\label{sec:setup}

\subsection{Overview of Initial Configuration}

For the main simulation run to be discussed, the initial magnetic field configuration is a force-free Harris sheet with added perturbation $\textbf{B}_p=B_{xp}\hat{\textbf{x}}+B_{zp}\hat{\textbf{z}}$:
\begin{equation}
\textbf{B}=B_0\left(\tanh(z/L)+B_{xp},1/\cosh(z/L),B_{zp}\right), 
\end{equation}

where $L$ is the current sheet half-width. The perturbation components have the form
\begin{eqnarray}
B_{xp}&=&-a_0x_m\frac{\pi}{2L_z}\exp\left(-\frac{x^2}{2x_m^2}+0.5\right)\sin\left(\frac{\pi z}{2L_z}\right)\nonumber\\
B_{zp}&=&a_0\frac{x}{x_m}\exp\left(-\frac{x^2}{2x_m^2}+0.5\right)\cos\left(\frac{\pi z}{2L_z}\right),\label{perturb}
\end{eqnarray}
where $L_z$ is the half-width of the numerical box in the $z$-direction, $a_0=0.1$ and $x_m=L_z/2$. This gives an X-point reconnection site at the centre of the numerical box, and allows the nonlinear phase of the evolution to be studied without considering the cause of the reconnection onset. 

The $x$-component of the force-free Harris sheet magnetic field (when $\textbf{B}_p=\textbf{0}$) has the same spatial structure as that of the Harris sheet \cite{Harris-1962}, and there is a non-uniform guide field in the $y$-direction, which is chosen in such a way that the total magnetic field strength is spatially uniform, and is given by  $B_0^2=B_x^2+B_y^2$. The resulting current density is parallel to the magnetic field, and hence the equilibrium is force-free \cite{Harrison-2009b,Neukirch-2009}. A further consequence is that both the plasma density and $P_{zz}$, the component of the pressure tensor that keeps the equilibrium in force balance, are spatially uniform. The equilibrium also has non-zero current density component in both the $x$- and $y$-directions, given by
\begin{eqnarray}
\textbf{j}=\frac{B_0}{\mu_0L}\frac{1}{\cosh^2(z/L)}\left(\sinh(z/L)\hat{\textbf{x}}+\hat{\textbf{y}}\right)\label{eqj}.
\end{eqnarray}


To initialise the particle positions and velocities in our main simulation run, we use the distribution function \cite{Harrison-2009b,Neukirch-2009}
\begin{eqnarray}
f_s&=&f_{0s}\exp(-\beta_sH_s)\nonumber\\
&{}&\times[\exp(\beta_su_{ys}p_{ys})+a_s\cos(\beta_su_{xs}p_{xs})+b_s],\label{ffhsdf}
\end{eqnarray}
where $H_s=(m_s/2)(v_x^2+v_y^2+v_z^2)$ is the particle energy, and $p_{xs}=m_sv_s+q_sA_x$ and $p_{ys}=m_sv_y+q_sA_y$ are the $x$- and $y$-components of the canonical momentum (for mass $m_s$, charge $q_s$ and vector potential components $A_x=2B_0L\arctan(e^{z/L})$ and $A_y=-B_0L\ln\left[\cosh(z/L)\right]$). The parameter $\beta_s$ is defined as $\beta_s=(k_BT_s)^{-1}$, where $T_s$ is the constant temperature of species $s$. Additionally, $f_{0s}$, $a_s$, $b_s$, $u_{xs}$ and $u_{ys}$ are constant parameters.

The distribution function (\ref{ffhsdf}) consists of a part which is equal to the Harris sheet distribution function \cite{Harris-1962}, and an extra part which arises from the non-uniform guide field of the force-free Harris sheet. It should be noted that it can have a non-Maxwellian structure in velocity space. For further details of the properties of this function, see Refs.~\onlinecite{Harrison-2009b} and ~\onlinecite{Neukirch-2009}. 

To analyse the expected transition from guide field to anti-parallel reconnection in the force-free Harris sheet case, we will also present results from two other simulation runs: one which starts from a Harris sheet, and the other from a Harris sheet plus uniform guide field of $B_y=B_0$.

\subsection{Normalisation and Parameters}

To study the reconnection process, we use a 2.5D fully electromagnetic particle-in-cell code, which has been frequently used by Hesse and co-authors (see, for example, Refs.~\onlinecite{Hesse-1999,Hesse-2001b}). The normalisation is as follows: the magnetic field is normalised to $B_0$; the number density to a free parameter, $n_0$; times to $\Omega_{i}^{-1}=(eB_0/m_i)^{-1}$ (the inverse of the ion cyclotron frequency in the equilibrium magnetic field); and lengths to the ion inertial length, $c/\omega_{i}$, where $\omega_i=(e^2n_0/\epsilon_0m_i)^{1/2}$ is the ion plasma frequency. Furthermore, velocities are normalised to the ion Alfv\'{e}n velocity, $v_A=\sqrt{\mu_0m_in_0}$, and so current densities and electric fields are normalised to $B_0/(\mu_0c/\omega_i)$ and $v_AB_0$, respectively.

In all simulation runs, we use an ion-electron mass ratio of $m_i/m_e=25$. The total number of particles is $1.44\times10^9$.
The grid spacing in $x$ and $z$ is $n_x=1200$, $n_z=600$, and hence there are $2000$ particles per cell. The numerical box has length $L_x=64.0$ and width $L_z=32.0$, which gives a grid spacing of $\Delta x=\Delta z=0.053$. The boundary conditions are periodic at the $x$-boundaries, and specularly reflecting at the $z$-boundaries. The time step chosen is $dt=0.5/\omega_{e}$ (where $\omega_e$ is the electron plasma frequency), with smaller time steps used occasionally. The ratio $\omega_{e}/\Omega_{e}$ is set to equal 5. The ion-electron temperature ratio is equal to unity, with $T_i+T_e=0.5$, so that $T_i=T_e=0.25$. The current sheet half-thickness is equal to one ion inertial length: $L=1.0$.


The various parameters from the force-free Harris sheet distribution function (\ref{ffhsdf}) have the following values: $u_{xe}/v_{th,e}=u_{ye}/v_{th,e}=\pm0.2$, $u_{xi}/v_{th,i}=u_{yi}/v_{th,i}=\pm1.0$, $a_e=0.52$, $a_i=1.36$, $b_e=1.02$ and $b_i=1.65$. Using conditions derived in Ref.~\onlinecite{Neukirch-2009}, it can be seen that this combination of parameters corresponds to a case where the ion distribution function is single-peaked in both $v_y$ and $v_z$, but has a double maximum in the $v_x$-direction, for small values of $z$ around zero. The electron distribution function is single-peaked in all three velocity components.




\section{Results}

\label{sec:results}

\subsection{Evolution of Magnetic Field and Current Density}

\begin{figure}[htp]
\scalebox{0.35}{\includegraphics[trim=2.3cm 1.1cm 3.4cm 2.2cm,clip=true]{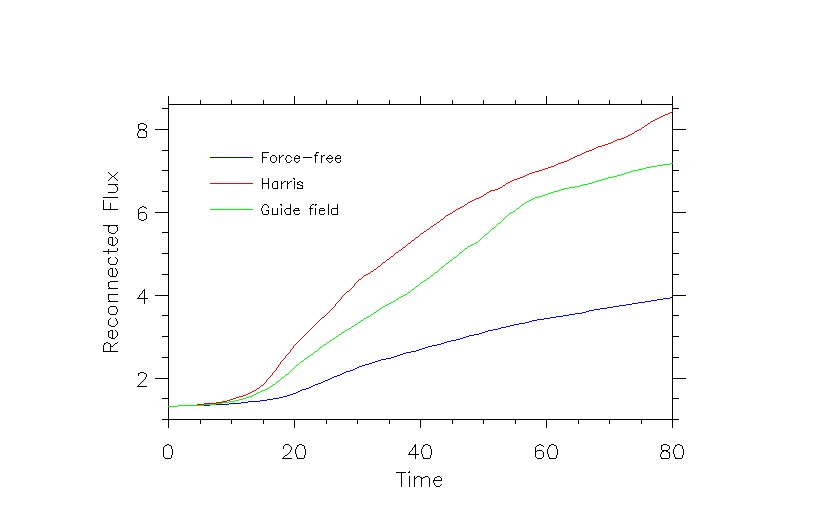}}
\caption{The reconnected flux for each simulation run.}\label{fig:recflux_compare}
\end{figure}

\begin{figure}[htp]
\scalebox{0.35}{\includegraphics[trim=2.3cm 1.1cm 3.4cm 2.2cm,clip=true]{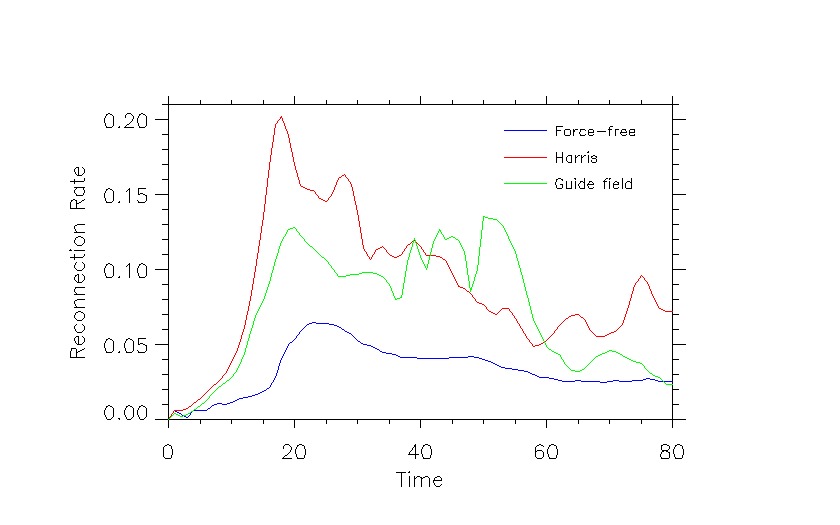}}
\caption{The reconnection rate for each simulation run.}\label{fig:recrate_compare}
\end{figure}



Figure \ref{fig:recflux_compare} shows the time evolution of the reconnected flux for the three simulation runs, and reconnection rates are shown in Figure \ref{fig:recrate_compare}. 
The maximum reconnection rate is highest in the Harris sheet case, occurring at $t=18$. It has been observed in previous work that the effect of a constant guide field (of significant magnitude) is to reduce the maximum reconnection rate \cite{Pritchett-2001,Ricci-2004,Hesse-2004}. We see here that in the force-free run the maximum reconnection rate is further reduced from that of the constant guide field. It should be noted, however, that we used a parameter combination such that the initial electron number density is 25\% higher in the force-free case than in the other two cases, which will have an effect on the reconnection rate. 
 

\begin{figure*}[htp]
\scalebox{0.3}{\includegraphics{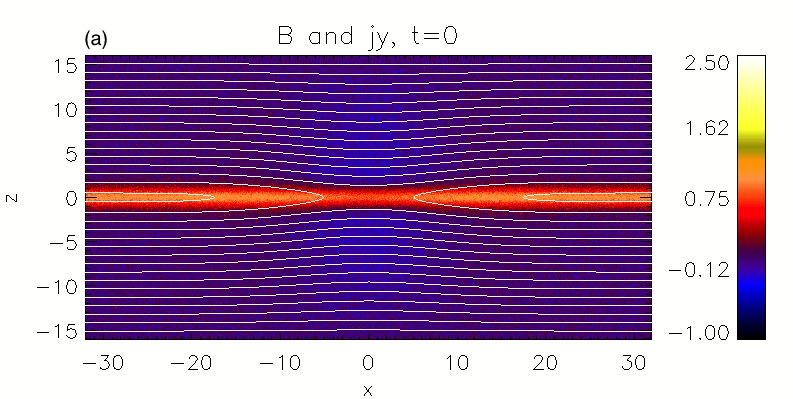}}\scalebox{0.3}{\includegraphics{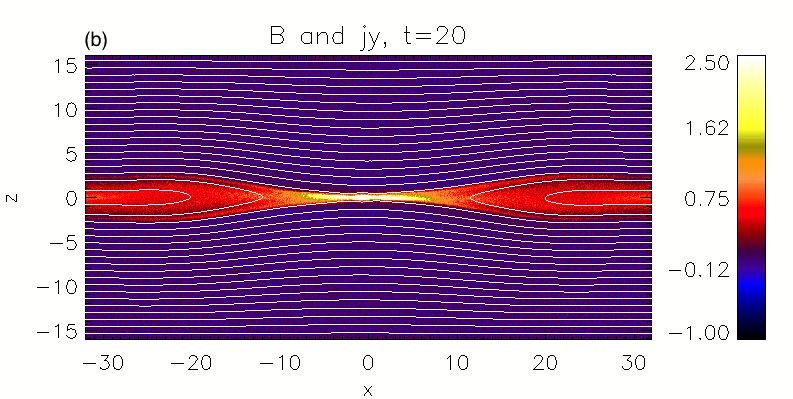}}\\
\scalebox{0.3}{\includegraphics{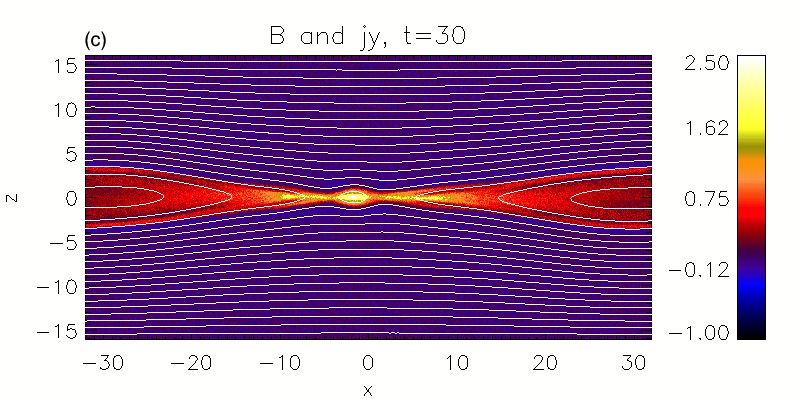}}\scalebox{0.3}{\includegraphics{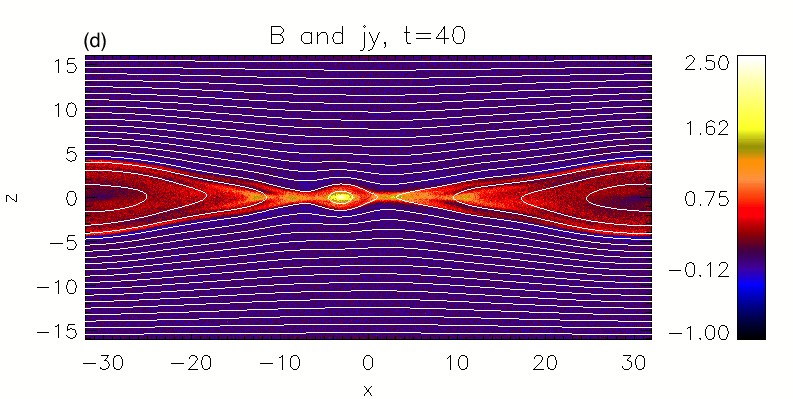}}\\\scalebox{0.3}{\includegraphics{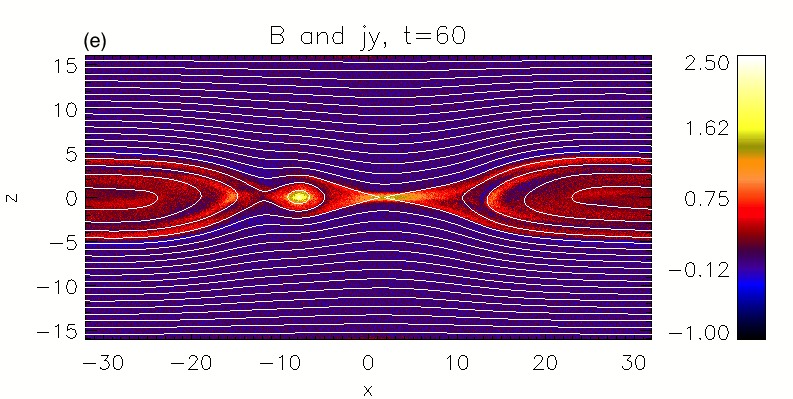}}\scalebox{0.3}{\includegraphics{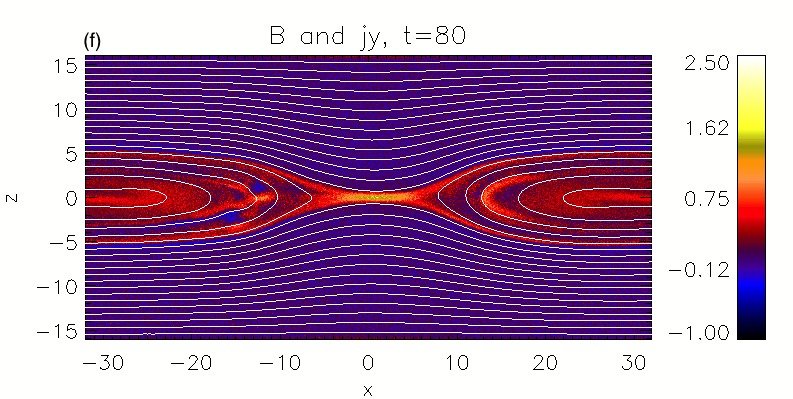}}
\caption{Evolution of $j_y$ (colour) and the magnetic field lines for the force-free run at (a) $t=0$, (b) $t=20$, (c) $t=30$, (d) $t=40$, (e) $t=60$, (f) $t=80$.}\label{fig:cxzjy_ffhs}
\end{figure*}



Figure \ref{fig:cxzjy_ffhs} shows the $y$-component of the current density (in colour) and the projection of the magnetic field lines on to the $x$-$z$-plane, at various times for the force-free run. The figures show how reconnection leads to global changes in the structure of both quantities. 
At $t=0$, it can be seen that the perturbation (\ref{perturb}) to the magnetic field gives an initial X-point in the centre of the box. As time proceeds initially, a strong current sheet develops in the central region, and is slightly inclined, which is a typical feature of guide field reconnection \cite{Hesse-2011}. As time proceeds beyond $t=20$, the current sheet becomes more aligned with the $x$-axis, which could be a sign of a transition from guide field to anti-parallel reconnection.

Looking closely at Figure \ref{fig:cxzjy_ffhs} for $t=20$, it can be seen that a small magnetic island has started to form, which is a result of the bifurcation of the original X-point reconnection site into two new reconnection regions \cite{Hesse-2001b} - one to the left of the island, and one to the right, which can be seen more clearly at later times. This is a feature commonly seen in reconnection simulations. 
Beyond $t=20$, the island proceeds to move to the left, and eventually disappears, as the right-hand X-point begins to dominate over the left-hand one. By the end of the simulation, at $t=80$, the island is no longer visible, and there is only one remaining reconnection region, which has shifted back towards the centre of the box. There is still a relatively strong current in this region though, which is higher than the original $j_y$.

\begin{figure*}[htp]
 \scalebox{0.35}{\includegraphics[trim=1.8cm 1.8cm 1.8cm 1.8cm,clip=true]{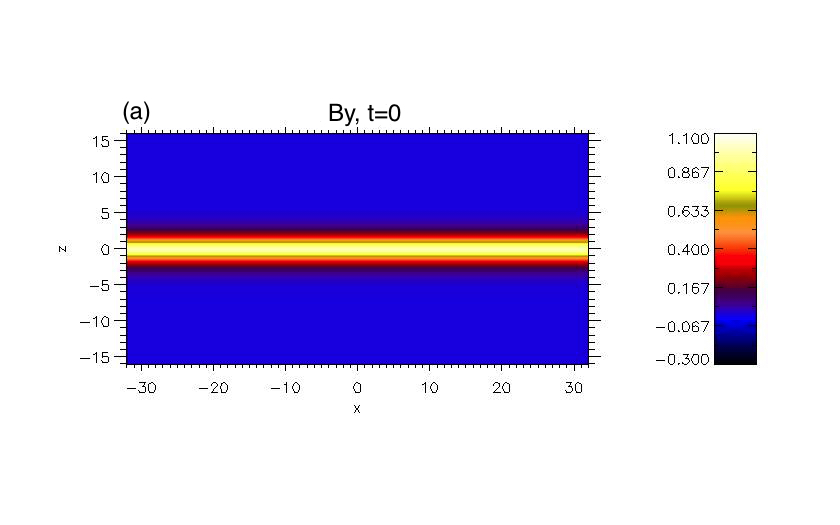}}\scalebox{0.35}{\includegraphics[trim=1.8cm 1.8cm 1.8cm 1.8cm,clip=true]{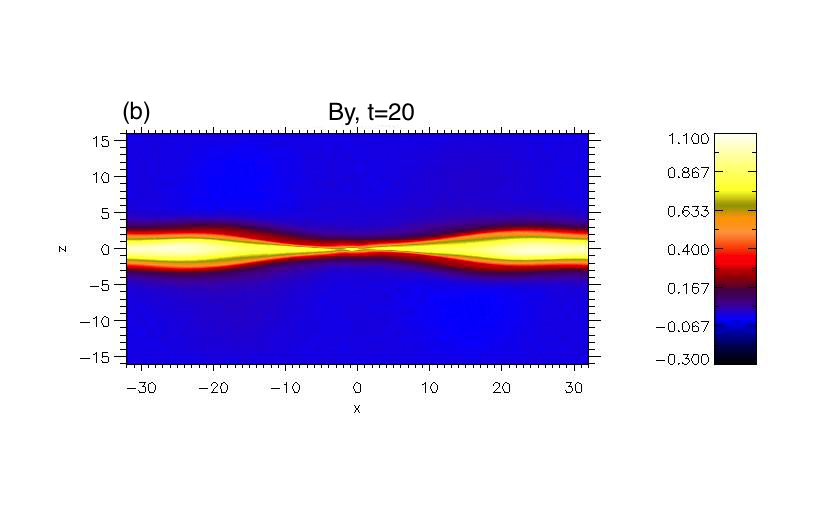}}\\
\scalebox{0.35}{\includegraphics[trim=1.8cm 1.8cm 1.8cm 1.8cm,clip=true]{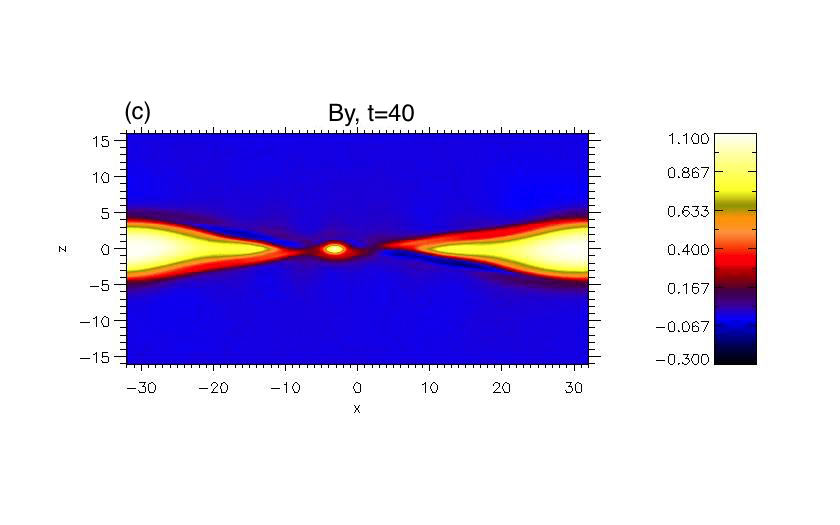}}\scalebox{0.35}{\includegraphics[trim=1.8cm 1.8cm 1.8cm 1.8cm,clip=true]{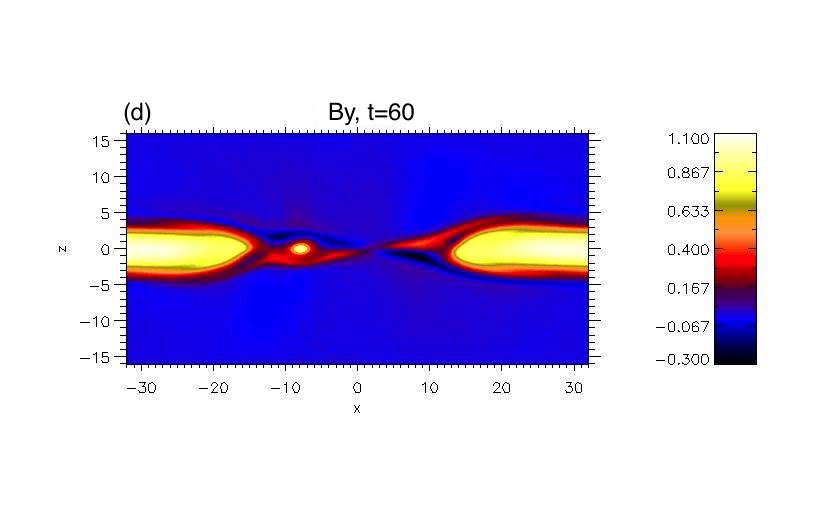}}
\caption{Evolution of $B_y$ in the force-free case at (a) $t=0$, (b) $t=20$, (c) $t=40$, (d) $t=60$.}\label{fig:by}
\end{figure*}

\begin{figure}[htp]
 \scalebox{0.35}{\includegraphics[trim=1.5cm 1.3cm 3.5cm 3.4cm,clip=true]{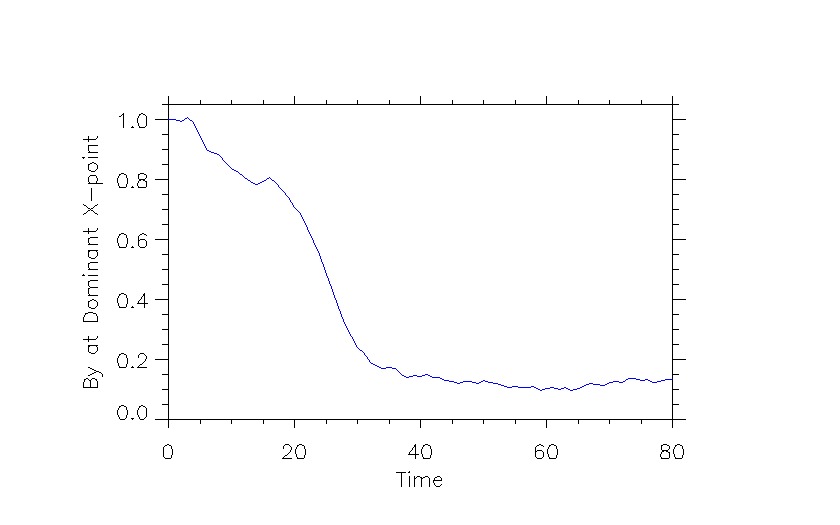}}
\caption{$B_y$ at the dominant X-point, as a function of time.}
\label{fig:byxpoint}
\end{figure}

Figure \ref{fig:by} shows the evolution of the non-uniform guide field in the force-free case. It can again be seen how the magnetic island starts to form around $t=20$, and eventually disappears. At $t=40$ and at subsequent times, a modified quadrupolar structure of $B_y$ can be seen around the X-point. This structure is qualitatively similar to that seen in Harris plus constant guide field simulations \cite{Pritchett-2001,Hesse-2002}, and so we do not see a transition to the quadrupolar structure seen in Harris sheet simulations \cite{Shay-1998,Hesse-2001b,Pritchett-2001}. 
Figure \ref{fig:byxpoint} shows the variation of $B_y$ at the dominant X-point with time. It can be seen that, on the whole, there is a downward trend as time proceeds, representing a gradual transition from guide field to anti-parallel reconnection (where $B_y$ would be close to zero). From around $t=35$ onwards, $B_y$ fluctuates around a value of approximately $0.15$. Of course, we do not have totally anti-parallel reconnection by the end of the simulation, but $B_y$ has clearly been significantly decreased from its initial value of 1.0 at the X-point.

\begin{figure*}[htp]
 \scalebox{0.3}{\includegraphics{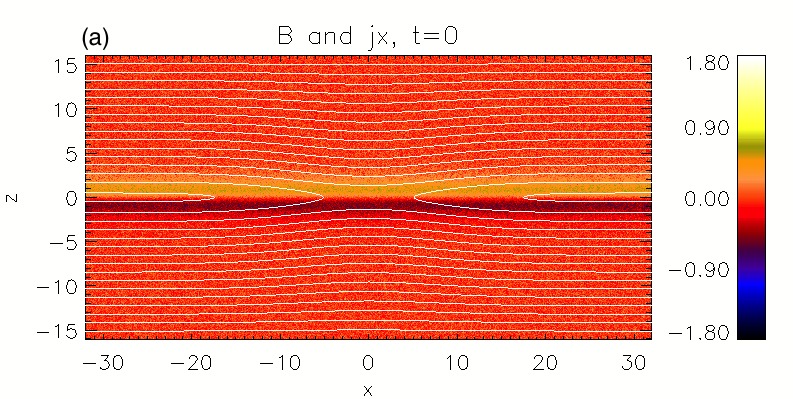}}\scalebox{0.3}{\includegraphics{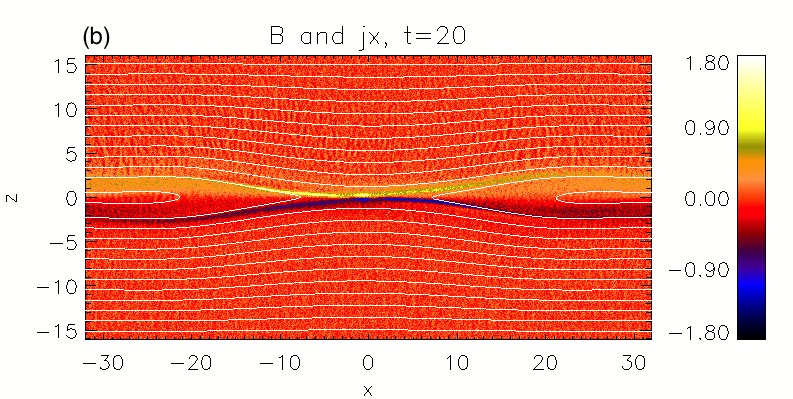}}
\\\scalebox{0.3}{\includegraphics{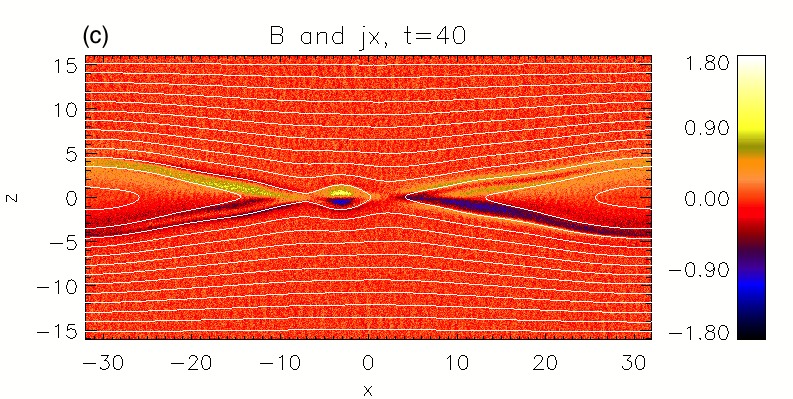}}\scalebox{0.3}{\includegraphics{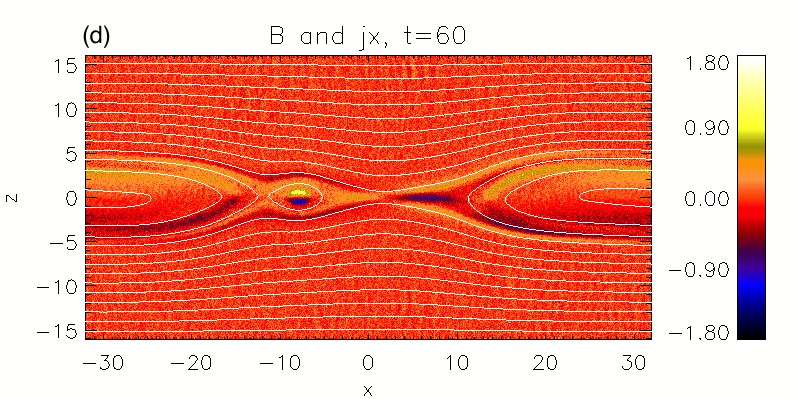}}
\caption{Evolution of $j_x$ (colour) and the magnetic field lines for the force-free run at (a) $t=0$, (b) $t=20$, (c) $t=40$, (d) $t=60$.}\label{fig:cxzjx}
\end{figure*}



Figure \ref{fig:cxzjx} shows the $x$-component of the current density in the force-free case. The equilibrium $j_x$ is anti-symmetric (see Eq (\ref{eqj})). As time proceeds, there is a build up of $j_x$ in the magnetic islands. These regions of strong $j_x$ correspond to regions where there is a strong gradient in the $y$-component of the magnetic field (see Figure \ref{fig:by}). Similar behaviour has been seen in linear force-free simulations, and also in preliminary force-free Harris sheet simulations \cite{mike-thesis}. We have not included similar plots for the Harris and Harris plus constant guide field runs, but comment that $j_x$ is more prominent in the force-free case, which is to be expected since the other two cases have zero equilibrium $j_x$.


\subsection{Electron Larmor Radius and Bounce Width}

\label{sec:rlebw}

In order to further investigate the expected transition from guide field to anti-parallel reconnection, we now consider the relevant length scales for the reconnection electric field. In the case of a guide field of significant magnitude, the electrons are strongly magnetised in the electron diffusion region, and $r_{Ley}={v_{th,e}}/{(eB_y/m_e)}$, the thermal electron Larmor radius in the guide field $B_y$,
is the characteristic length scale \cite{Hesse-2011}. As the guide field gets weaker, however, 
 the important scale length is the electron bounce width in the reconnecting field component $B_x$, given by
\begin{equation}
\lambda_z=\left(\frac{2m_ek_BT_e}{e^2(\partial B_x/\partial z)^2}\right)^{1/4}.\label{lambdaz}
\end{equation}

As discussed in Ref.~\onlinecite{Hesse-2011}, the effect of the guide field $B_y$ on the electron orbits is significant if 
\begin{equation}
r_{Ley}\le\lambda_z\label{ineq}.
\end{equation}
When the condition (\ref{ineq}) is satisfied at the reconnection site, therefore, we would expect to see mainly signatures of guide field reconnection and, when it is no longer satisfied, we would expect that this has coincided with a gradual transition towards anti-parallel reconnection, and would expect to see some signatures of this.

\begin{figure}[htp]
\scalebox{0.34}{\includegraphics[trim=2.05cm 0.8cm 2.8cm 2.8cm,clip=true]{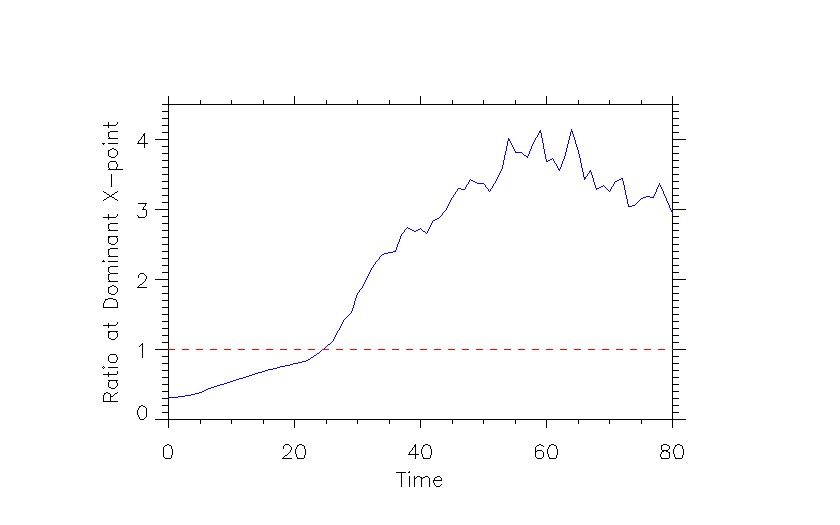}}
\caption{\footnotesize{Ratio of the electron Larmor radius in the guide field $B_y$ and the electron bounce width, $\lambda_z$, plotted against time. A horizontal line is plotted at $r_{Ley}/\lambda_z=1$.}}\label{fig:rlebw_time}
\end{figure}

In Figure \ref{fig:rlebw_time}, the ratio $r_{Ley}/\lambda_z$ is plotted as a function of time. It first goes above unity between $t=24$ and $t=25$. We will consider $t=25$ to be the 'transition time' towards anti-parallel reconnection, since after this time the ratio ceases to fluctuate around unity.
\begin{figure}[htp]
\scalebox{0.32}{\includegraphics{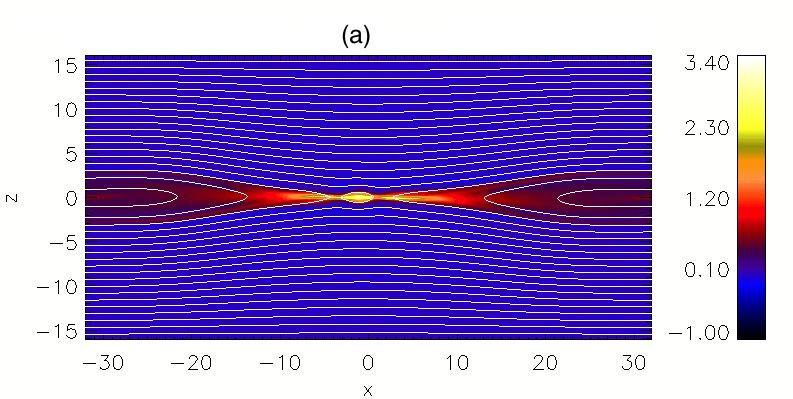}}\\\scalebox{0.32}{\includegraphics{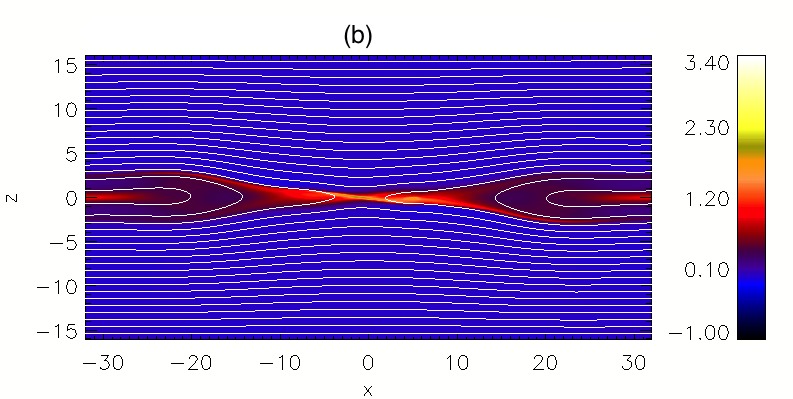}}\\
\scalebox{0.32}{\includegraphics{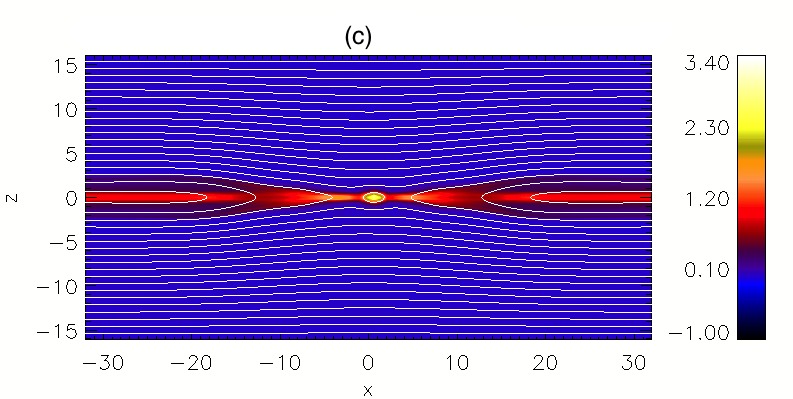}}
\caption{Evolution of $j_y$ and the magnetic field lines, for (a) the force-free case at the 'time of transition' to anti-parallel reconnection (t=25), and the corresponding times at which the reconnected flux is the same in (b) the constant guide field case (t=18) and (c) the Harris case (t=16).}\label{fig:cxzjy_transition}
\end{figure}
Figure \ref{fig:cxzjy_transition} shows the $y$-component of the current density and the magnetic field lines at this time (for the force-free run), together with plots at $t=18$ and $t=16$ for the Harris plus constant guide field and Harris runs, respectively (these are the times at which the reconnected flux in both cases matches that of the force-free case at $t=25$). On the macroscopic level, the field-line structure looks more like that from the Harris sheet case, with an island separating two X-points. The central current sheet in the force-free case is still slightly inclined, but not as much as seen in Figure \ref{fig:cxzjy_ffhs} at $t=20$, and this inclination is also not as strong as seen in the constant guide field case.

\subsection{The Reconnection Electric Field} 

\label{sec:ey}

In a 2D setup, the reconnection electric field is given by
\begin{eqnarray}
E_y&=&(v_{xe}B_z-v_{ze}B_x)-\frac{1}{en_e}\left(\frac{\partial P_{xye}}{\partial x}+\frac{\partial P_{yze}}{\partial z}\right)\nonumber\\
&{}&-\frac{m_e}{e}\left(\frac{\partial v_{ye}}{\partial t}+v_{ex}\frac{\partial v_{ey}}{\partial x}+v_{ez}\frac{\partial v_{ey}}{\partial z}\right),\label{eyc}
\end{eqnarray}
where the first bracket represents convection, the second represents the effect of the off-diagonal pressure tensor components, and the last bracket represents the effect of bulk inertia. 

\begin{figure}[htp]
\scalebox{0.33}{\includegraphics[trim=1.8cm 1.8cm 1.8cm 1.8cm,clip=true]{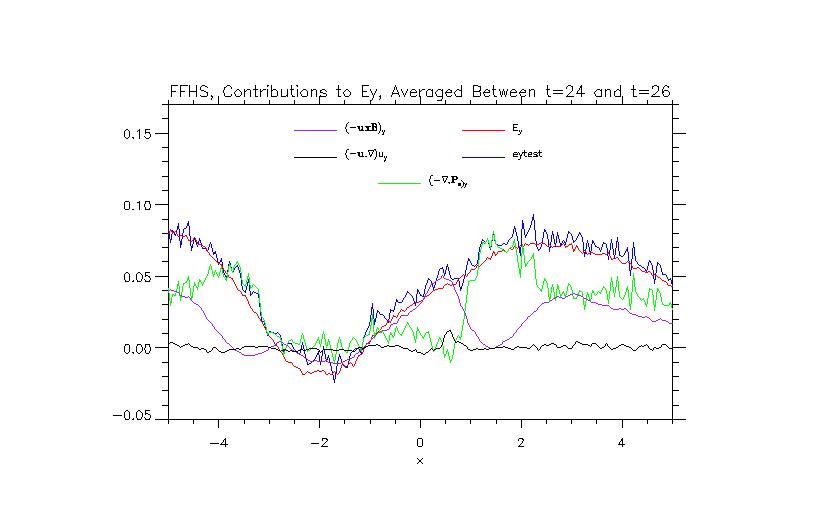}}
\caption{Contributions to $E_y$ along $x$, through the dominant X-point, for data averaged around the transition point at $t=25$.}\label{fig:ey_cont_transition_x}
\end{figure}

\begin{figure}[htp]
\scalebox{0.33}{\includegraphics[trim=1.8cm 1.8cm 1.8cm 1.8cm,clip=true]{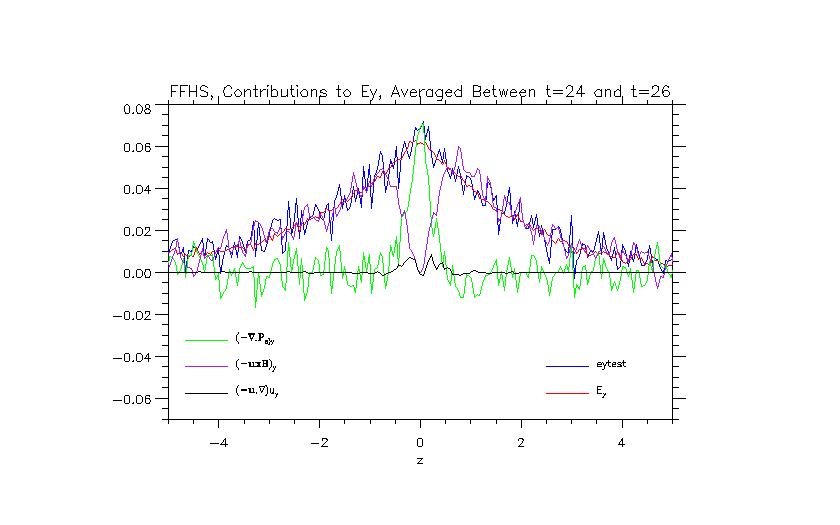}}
\caption{Contributions to $E_y$ along $z$, through the dominant X-point, for data averaged around the transition point at $t=25$.}\label{fig:ey_cont_transition_z}
\end{figure}

Figures \ref{fig:ey_cont_transition_x} and \ref{fig:ey_cont_transition_z} show, for the force-free case, the contributions from each of the terms on the right-hand-side of Eq. (\ref{eyc}) to the reconnection electric field, along $x$ and $z$, through the average position of the dominant X-point, for data averaged between $t=24$ and $t=26$. The time we chose to average around is the 'transition time' discussed in Section \ref{sec:rlebw}, where the dominant scale for the evolution switches from the Larmor radius in the guide field $B_y$, to the electron bounce width $\lambda_z$. The pressure gradient terms are graphed as green lines, the convection term as purple lines, and the inertial term as black lines. The sum of these three terms, referred to as 'eytest', is plotted as a blue line in both plots. Although this fluctuates due to random noise, it can be seen that in both plots it matches reasonably well with the $E_y$ that is calculated on the numerical grid in the code (indicated by red lines).

From Figure \ref{fig:ey_cont_transition_x}, it can be seen that the pressure gradient term in $x$ is significantly enhanced around the dominant X-point ($x=1.39$). This increase in pressure coincides with a decrease (towards zero) of the convection term. Such behaviour can also be seen at $x\approx-3.75$, which corresponds roughly to the position of the second, less dominant X-point (see Figure \ref{fig:cxzjy_transition}). In comparison with the other terms, the inertial term is small. The convection term should of course vanish at any X-points, since they are stagnation points where $\textbf{v}_s=0$, and so the pressure gradient term acts to support the reconnection electric field. This is in agreement with what has been found previously for Harris sheet and Harris sheet plus constant guide field simulations \cite{Hesse-1999,Kuznetsova-1998,Kuznetsova-2000,Kuznetsova-2001,Pritchett-2001,Hesse-2004,Ricci-2004,Pritchett-2005}.


From Figure \ref{fig:ey_cont_transition_z}, it can be seen that, at $z=0$ (the position of the dominant X-point), the convection term can be seen to drop to zero, and again the main contribution to $E_y$ comes from the pressure term. The inertial term is close virtually zero everywhere, apart from in the small region surrounding the X-point.

\subsection{Pressure Tensor Components}

\label{sec:pressure}

We now focus on the structure of the off-diagonal components of the electron pressure tensor in the diffusion region, restricting attention to the electron quantities, since they are the dominant current carriers. 
Of particular importance are the non-gyrotropic components, which are given by \cite{Hesse-2011}
\begin{equation}
\textbf{\underline{P}}_{e,ng}=\textbf{\underline{P}}_e-\textbf{\underline{P}}_{e,g},
\end{equation}
where
\begin{equation}
\textbf{\underline{P}}_{e,g}=p_{\perp}\textbf{\underline{I}}+\frac{p_{\parallel}-p_\perp}{B^2}\textbf{B}\textbf{B}
\end{equation}
is the gyrotropic component. The term $(\nabla\cdot\textbf{\underline{P}}_{e,g})_y$ vanishes at any X-points, since $B_x$ and $B_z$ vanish, and so non-gyrotropies of the pressure are required to give a contribution to the reconnection electric field \cite{Hesse-2004}.

\begin{figure*}[htp]
\scalebox{0.32}{\includegraphics[trim=2.0cm 3.4cm 1.8cm 1.8cm,clip=true]{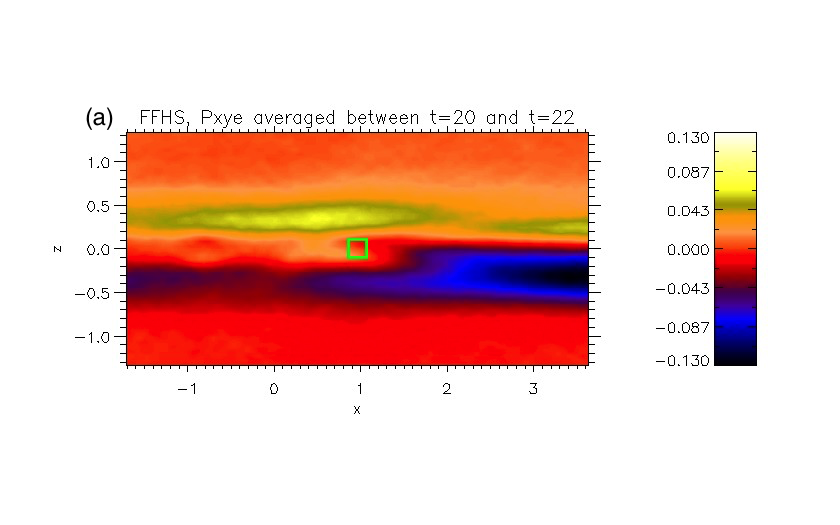}}\scalebox{0.32}{\includegraphics[trim=2.0cm 3.4cm 1.8cm 1.8cm,clip=true]{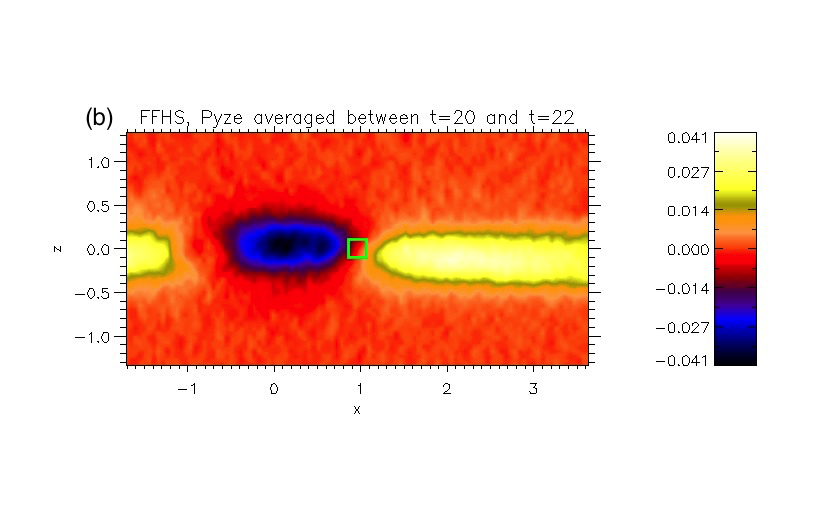}}\\
\scalebox{0.32}{\includegraphics[trim=2.0cm 3.4cm 1.8cm 1.8cm,clip=true]{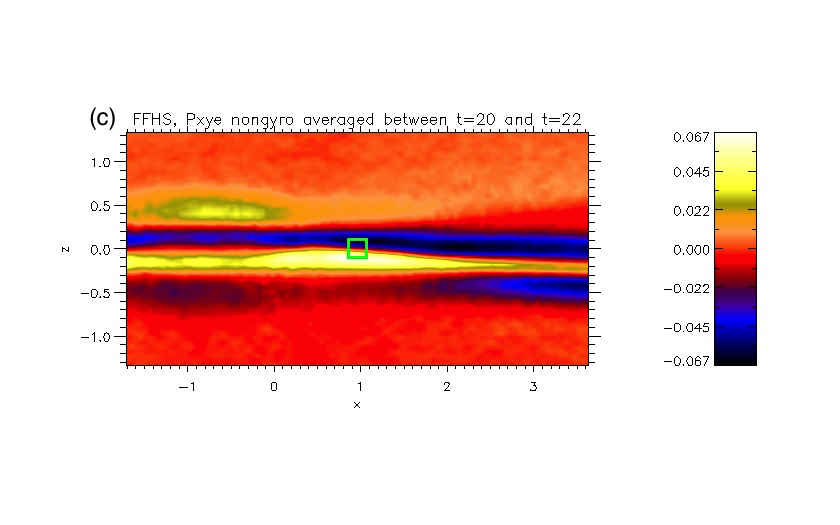}}\scalebox{0.32}{\includegraphics[trim=2.0cm 3.4cm 1.8cm 1.8cm,clip=true]{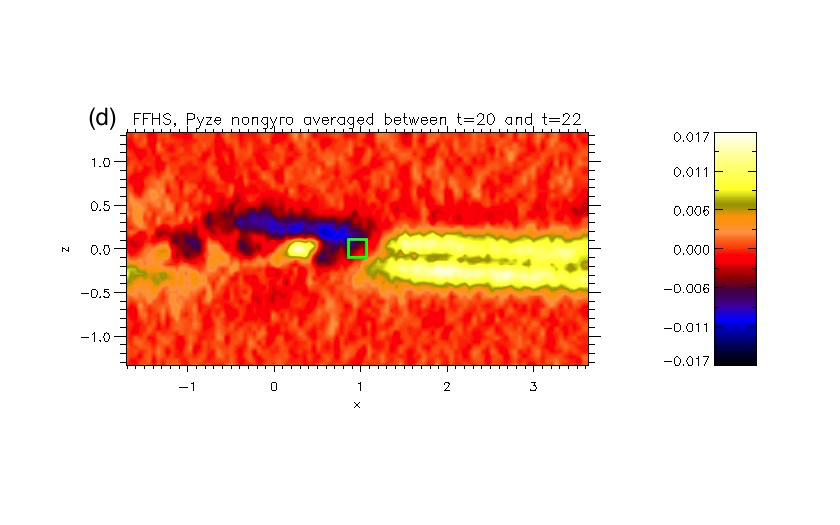}}      
\caption{\footnotesize{Electron pressure tensor components for the force-free case, for data averaged between $t=20$ and $t=22$. Shown are (a) $P_{xye}$, (b) $P_{yze}$, (c) $P_{xye,ng}$, (d) $P_{yze,ng}$.}}
\label{fig:p_early_ffhs}
\end{figure*}

\begin{figure*}[htp]
\scalebox{0.32}{\includegraphics[trim=2.0cm 3.4cm 1.8cm 1.8cm,clip=true]{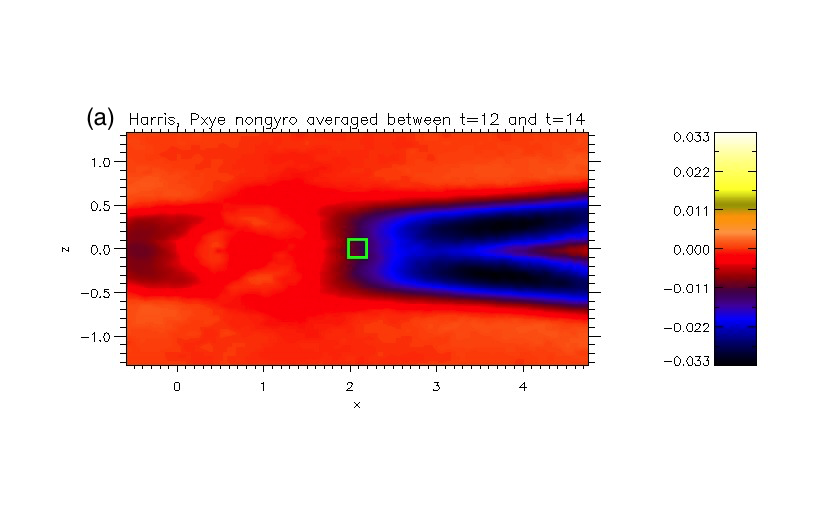}}\scalebox{0.32}{\includegraphics[trim=2.0cm 3.4cm 1.8cm 1.8cm,clip=true]{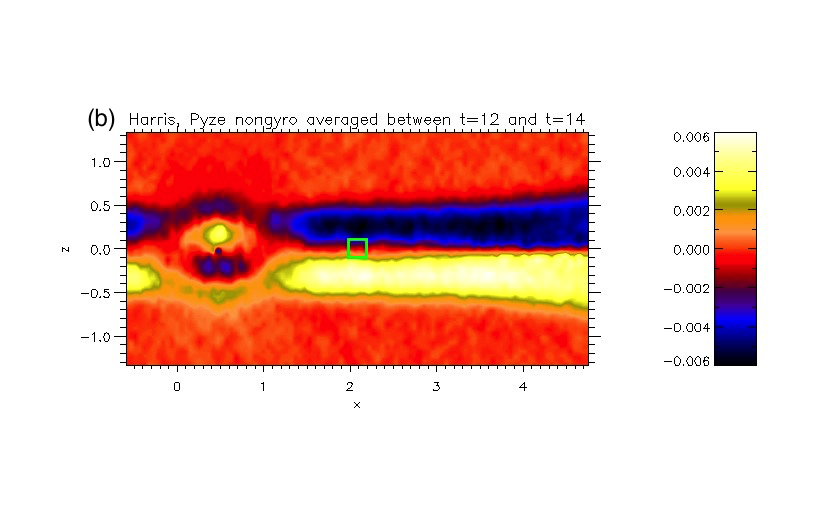}}
\caption{\footnotesize{
Electron pressure tensor components for the Harris case, for data averaged between $t=12$ and $t=14$. Shown are (a) $P_{xye,ng}$, (b) $P_{yze,ng}$.}}
\label{fig:p_early_harris}
\end{figure*}

\begin{figure*}[htp]
\scalebox{0.32}{\includegraphics[trim=2.0cm 3.4cm 1.8cm 1.8cm,clip=true]{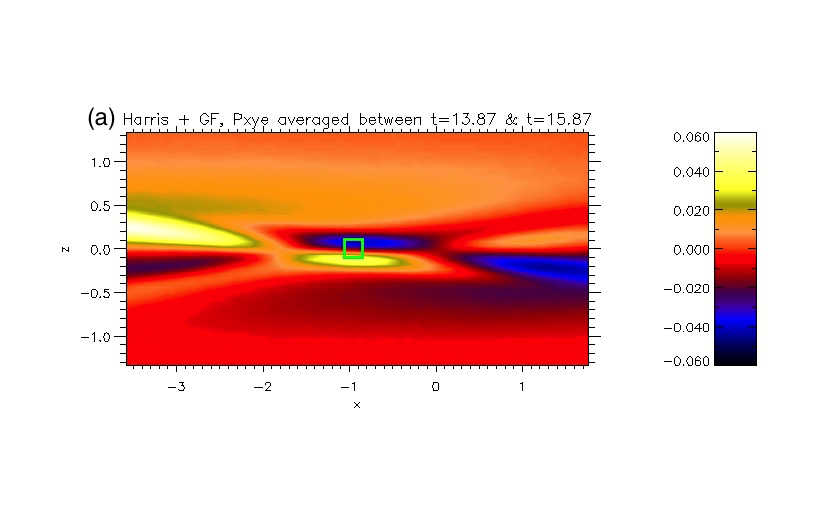}}\scalebox{0.32}{\includegraphics[trim=2.0cm 3.4cm 1.8cm 1.8cm,clip=true]{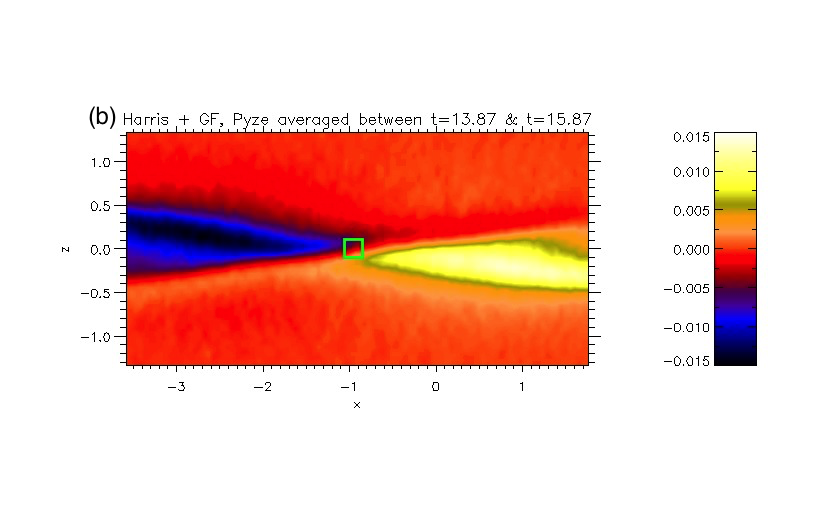}}\\
\scalebox{0.32}{\includegraphics[trim=2.0cm 3.4cm 1.8cm 1.8cm,clip=true]{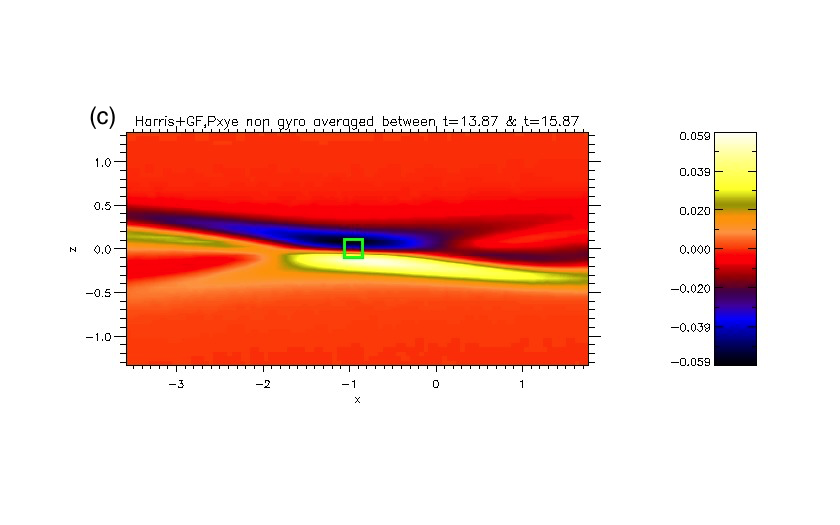}}\scalebox{0.32}{\includegraphics[trim=2.0cm 3.4cm 1.8cm 1.8cm,clip=true]{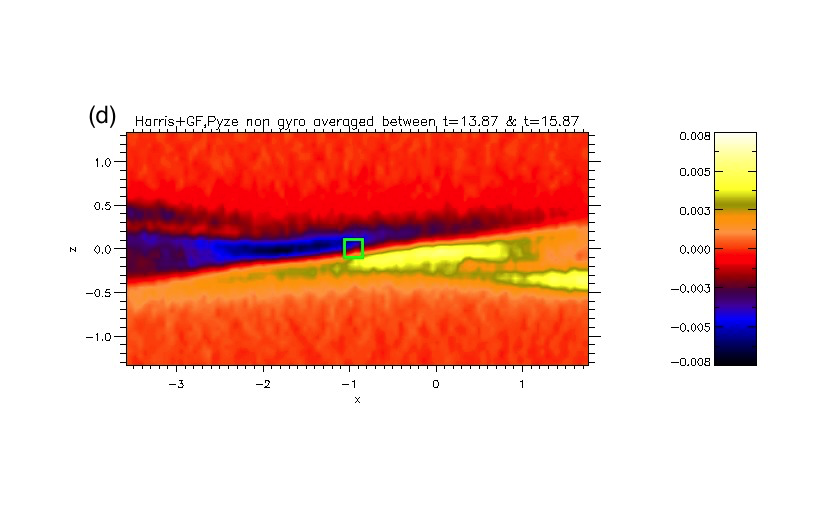}}    
\caption{\footnotesize{Electron pressure tensor components for the constant guide field case, for data averaged between $t=13.88$ and $t=15.88$. Shown are (a) $P_{xye}$, (b) $P_{yze}$, (c) $P_{xye,ng}$, (d) $P_{yze,ng}$.}}
\label{fig:p_early_gf}
\end{figure*}

Figures \ref{fig:p_early_ffhs} to \ref{fig:p_early_gf} show plots of the $xy$- and $yz$-components of the electron pressure tensor, together with the corresponding non-gyrotropic parts, at an early stage of the evolution, at which the total reconnected flux is the same in each case. The data has been averaged between $t=20$ and $t=22$ for the force-free run, $t=12$ and $t=14$ for the Harris run, and $t=13.87$ and $t=15.87$ for the constant guide field run. Note that we only show the non-gyrotropic components in the Harris case (Figure \ref{fig:p_early_harris}) because they are almost identical to the plots of the total $P_{xye}$ and $P_{yze}$. The average location of the X-point under consideration is indicated by a green square. From Figure \ref{fig:p_early_ffhs} for the force-free case, it can be seen that both $P_{xye}$ and $P_{xye,ng}$ have a gradient primarily in $z$, which is comparable to that from the constant guide field case in Figure \ref{fig:p_early_gf}. The structure of $P_{yze}$ and $P_{yze,ng}$ are also comparable to that from the constant guide field case in the vicinity of the X-point - these components all have gradients in $x$. Note, however, that $P_{yze,ng}$ in the constant guide field case also has a significant gradient in $z$, and so there is a significant difference in this component between the force-free and constant guide field cases. The structure of all pressure components in the vicinity of the X-point for the force-free case differ considerably from that of the Harris sheet case, which clearly has horizontal gradients in $P_{xye}$ and vertical gradients in $P_{yze}$.
\begin{figure*}[htp]
\scalebox{0.32}{\includegraphics[trim=2.0cm 3.4cm 1.8cm 1.8cm,clip=true]{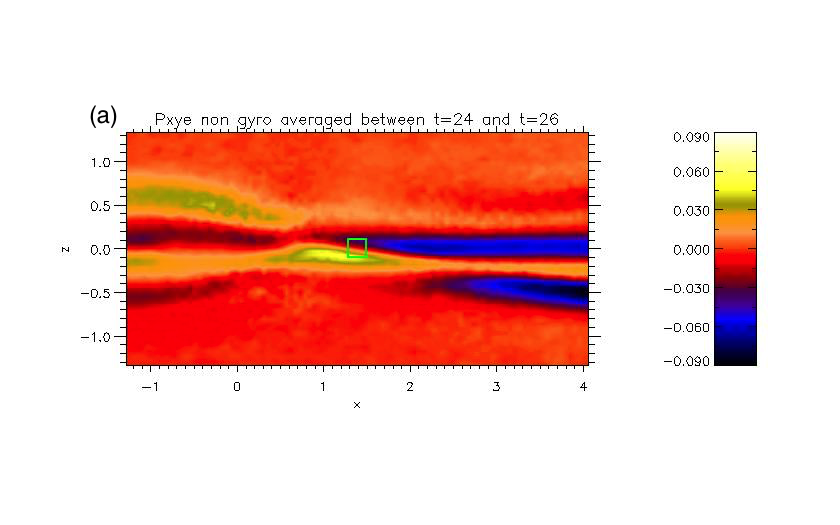}}\scalebox{0.32}{\includegraphics[trim=2.0cm 3.4cm 1.8cm 1.8cm,clip=true]{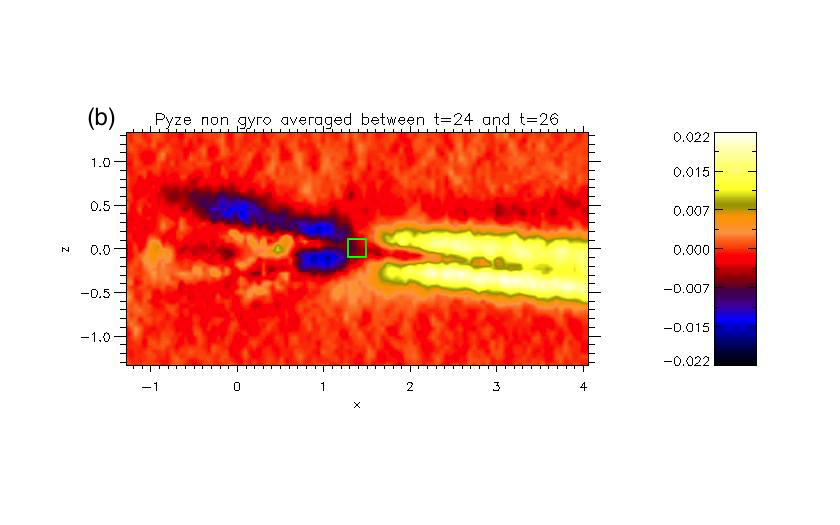}}
\caption{\footnotesize{Electron pressure tensor components for the force-free case, for data averaged between $t=24$ and $t=26$ (around the 'transition time' at $t=25$). Shown are (a) $P_{xye,ng}$, (b) $P_{yze,ng}$.}}
\label{fig:p_transition_ffhs}
\end{figure*}
 It can be said, therefore, that in the early stages of the evolution the pressure in the force-free case exhibits (qualitatively) more features of guide field reconnection than anti-parallel reconnection.
 
 As we discussed in Section \ref{sec:rlebw}, there is a change in the important scale length for the evolution around $t=25$, from the Larmor radius in the guide field $B_y$, to the electron bounce width $\lambda_z$. Figure \ref{fig:p_transition_ffhs} shows non-gyrotropic pressure plots for the force-free case, for data averaged around this transition time (between $t=24$ and $t=26$). On the whole, in the vicinity of the X-point, the structures remain qualitatively more similar to those from the constant guide field case (Figure \ref{fig:p_early_gf}), than the Harris case (Figure \ref{fig:p_early_harris}).



\begin{figure*}[htp]
\scalebox{0.32}{\includegraphics[trim=2.0cm 3.4cm 1.8cm 1.8cm,clip=true]{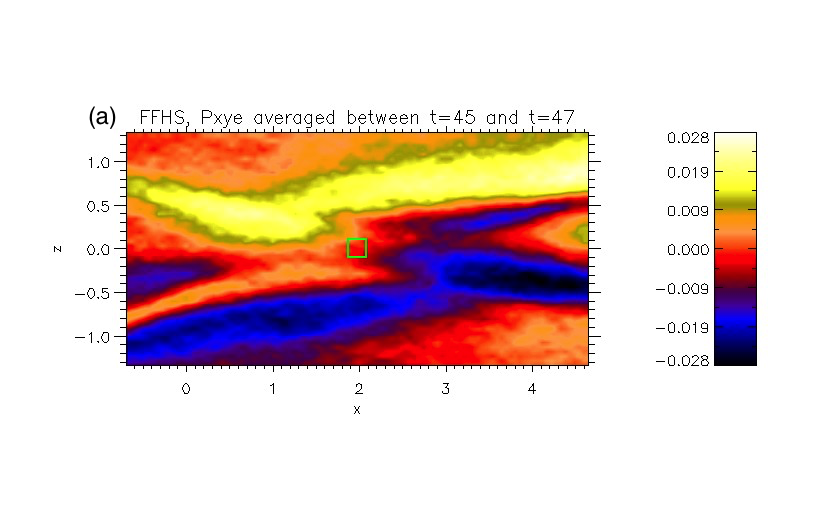}}\scalebox{0.32}{\includegraphics[trim=2.0cm 3.4cm 1.8cm 1.8cm,clip=true]{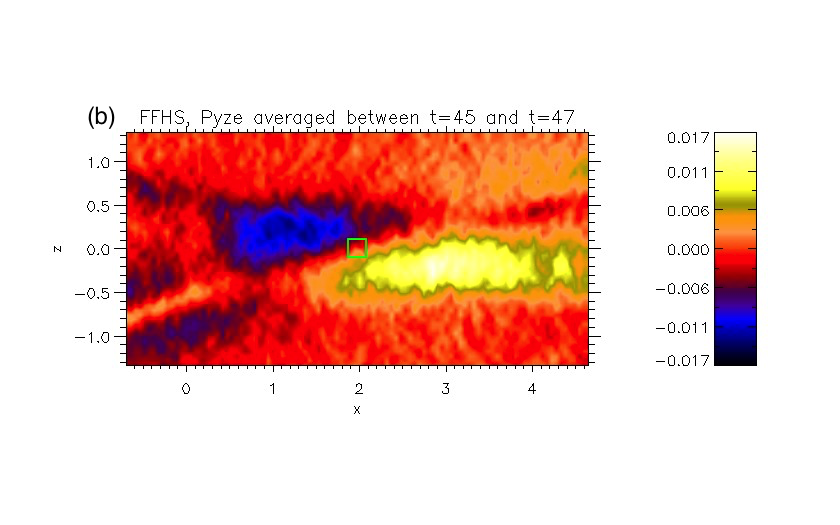}}\\
\scalebox{0.32}{\includegraphics[trim=2.0cm 3.4cm 1.8cm 1.8cm,clip=true]{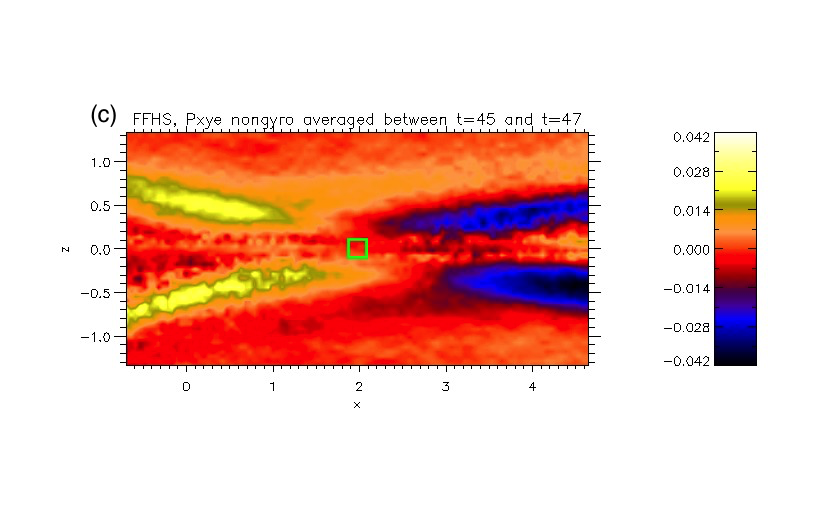}}\scalebox{0.32}{\includegraphics[trim=2.0cm 3.4cm 1.8cm 1.8cm,clip=true]{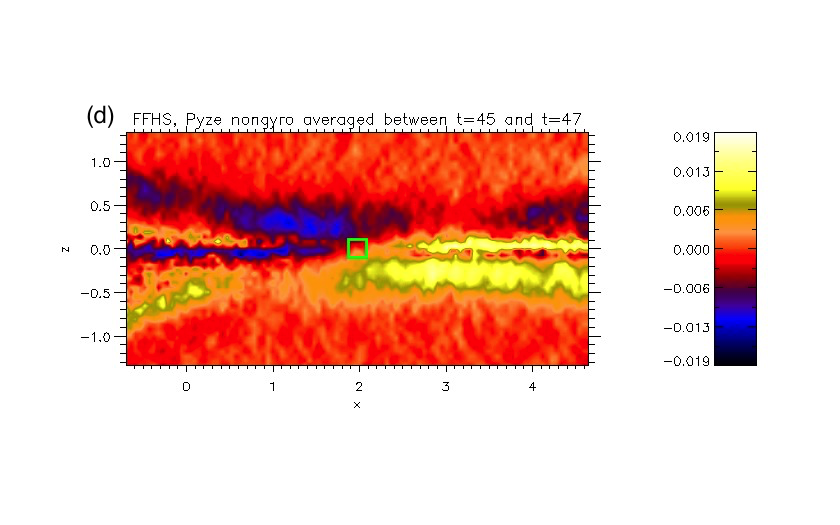}}
\caption{\footnotesize{Electron pressure tensor components for the force-free case, for data averaged between $t=45$ and $t=47$. Shown are (a) $P_{xye}$, (b) $P_{yze}$, (c) $P_{xye,ng}$, (d) $P_{yze,ng}$.}}
\label{fig:p_later_ffhs}
\end{figure*}

\begin{figure*}[htp]
\scalebox{0.32}{\includegraphics[trim=2.0cm 3.4cm 1.8cm 1.8cm,clip=true]{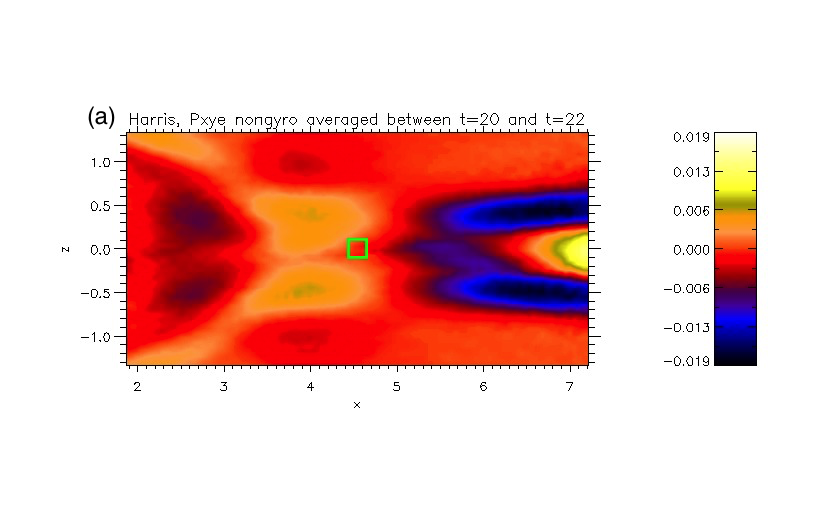}}\scalebox{0.32}{\includegraphics[trim=2.0cm 3.4cm 1.8cm 1.8cm,clip=true]{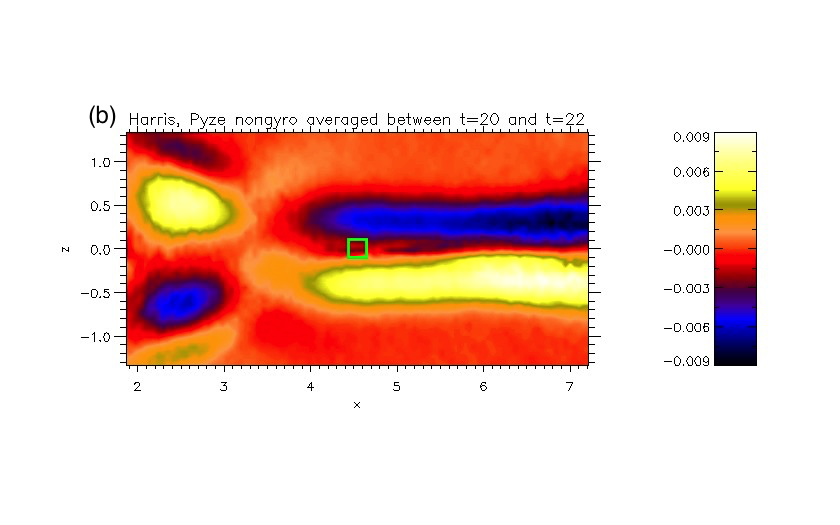}}
\caption{\footnotesize{
Electron pressure tensor components for the Harris case, for data averaged between $t=25$ and $t=27$. Shown are (a) $P_{xye,ng}$, (b) $P_{yze,ng}$.}}
\label{fig:p_later_harris}
\end{figure*}

\begin{figure*}[htp]
\scalebox{0.32}{\includegraphics[trim=2.0cm 3.4cm 1.8cm 1.8cm,clip=true]{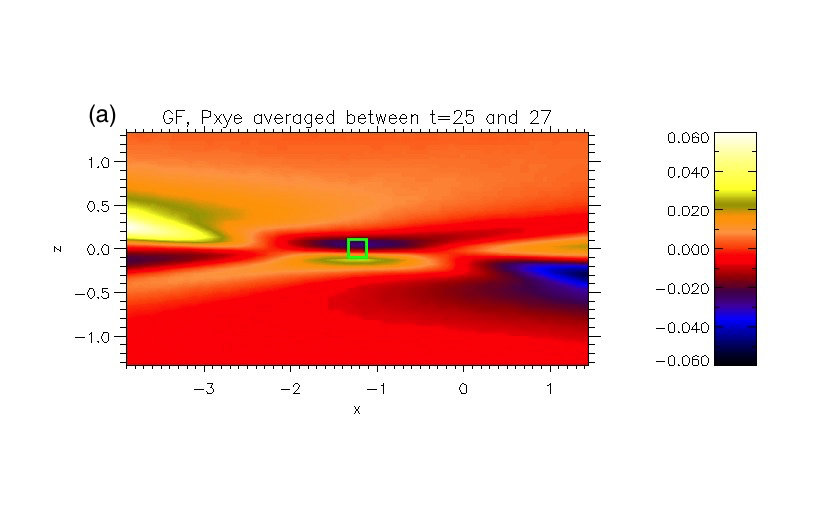}}\scalebox{0.32}{\includegraphics[trim=2.0cm 3.4cm 1.8cm 1.8cm,clip=true]{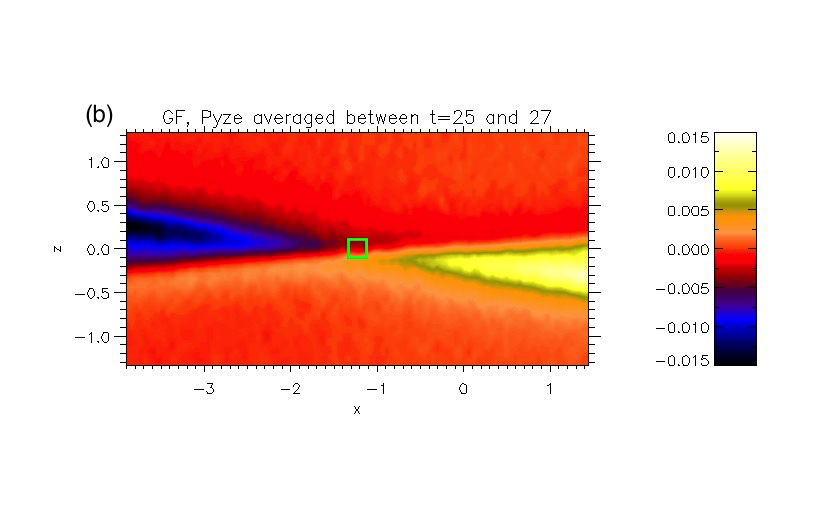}}\\
\scalebox{0.32}{\includegraphics[trim=2.0cm 3.4cm 1.8cm 1.8cm,clip=true]{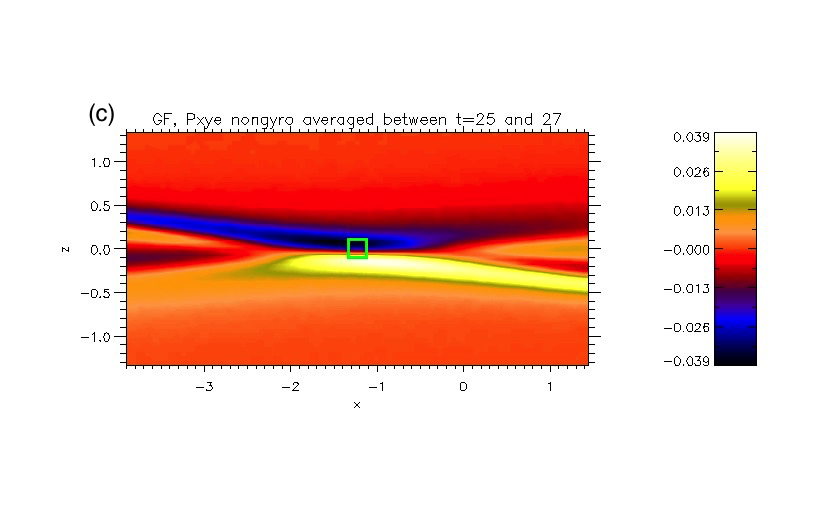}}\scalebox{0.32}{\includegraphics[trim=2.0cm 3.4cm 1.8cm 1.8cm,clip=true]{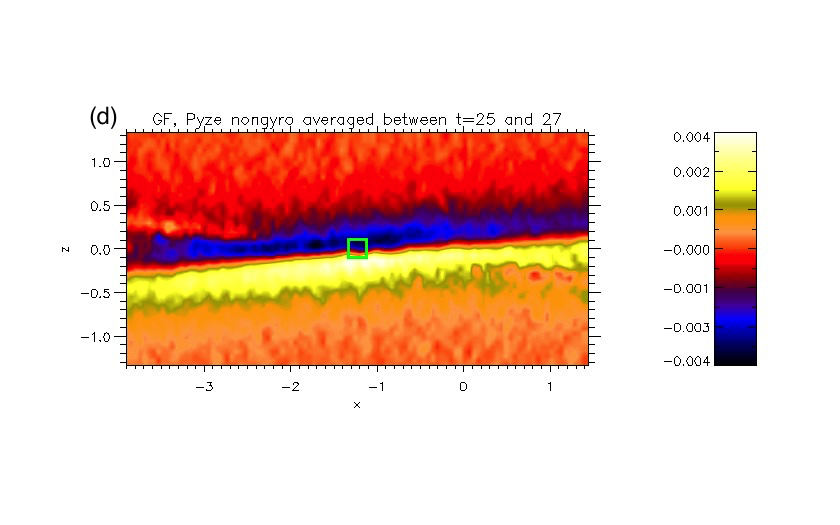}}
\caption{\footnotesize{Electron pressure tensor components for the Harris plus constant guide field case, for data averaged between $t=25$ and $t=27$. Shown are (a) $P_{xye}$, (b) $P_{yze}$, (c) $P_{xye,ng}$, (d) $P_{yze,ng}$.}}
\label{fig:p_later_gf}
\end{figure*}

To further investigate the transition, therefore, we now focus on a later time in the evolution. Figures \ref{fig:p_later_ffhs} to \ref{fig:p_later_gf} show the pressure components for data averaged between $t=45$ and $t=47$ for the force-free case, $t=25$ and $t=27$ for the constant guide field case, and $t=20$ and $t=22$ for the Harris case. As with the earlier Harris plot (Figure \ref{fig:p_early_harris}), we only show the non-gyrotropic components in Figure \ref{fig:p_later_harris} because again they are almost identical to the plots of the total $P_{xye}$ and $P_{yze}$. The structure of both $P_{xye}$ and $P_{xye,ng}$ in the force-free case is now significantly different than at earlier times. Focussing on the non-gyrotropic component, $P_{xye,ng}$, the gradient is now primarily in the horizontal direction, and looks comparable (qualitatively) to $P_{xye,ng}$ for the Harris sheet. The other non-gyrotropic component, $P_{yze,ng}$, now has significant gradients in both $x$ and $z$, and still looks more similar to $P_{yze}$ in the guide field case than in the Harris sheet case. From Figures \ref{fig:p_later_ffhs} to \ref{fig:p_later_gf}, we can conclude that some sort of transition has taken place in the structure of the pressure, since we see some signatures of anti-parallel reconnection. We can also conclude from this that the transition is not as simple as being from purely guide field reconnection to purely anti-parallel reconnection, but instead we see initially primarily signatures of guide field reconnection, and signatures of both guide field and anti-parallel reconnection as the system evolves. This may be due to the fact that while $B_y$ at the dominant reconnection site (see Figure \ref{fig:byxpoint}) decreases over time, it does not actually vanish completely, and Figure \ref{fig:by} shows that there is a modified quadrupolar structure of $B_y$ at later times - so not a transition to the quadrupolar structure seen in Harris sheet simulations. We speculate that this could cause some features of guide field reconnection to persist. This is clearly a point which should be investigated in future studies.

\section{Summary and Conclusions}
\label{sec:summary}

In this paper, we have investigated how the reconnection process differs when adding a non-uniform guide field to the Harris sheet, instead of a constant guide field. We have presented results from a 2.5D fully electromagnetic particle-in-cell simulation of collisions magnetic reconnection, starting from a force-free Harris sheet with added perturbation and using the exact collisionless distribution function solution from Ref.~\onlinecite{Harrison-2009b} to initialise the particle velocities. For comparison, we have also presented results from a Harris sheet simulation, and a Harris sheet plus uniform guide field simulation.


We have found, as expected, that as time evolves in the force-free Harris sheet simulation, there are signs of a transition from guide field to anti-parallel reconnection. Firstly, on the macroscopic level, the initially rotated current sheet (similar to the constant guide field case) becomes more horizontally oriented (more like the Harris sheet case) as time progresses. Secondly, there is a gradual decrease in the guide field $B_y$ at the dominant X-point, indicating that it becomes less important as time proceeds. Thirdly, the transition can also be seen by looking at the ratio of the electron Larmor radius in the guide field $B_y$, and the electron bounce width in the reconnecting field component, $B_x$. The effect of the guide field on the electron orbits is significant if the ratio is less than unity \cite{Hesse-2011}. At the beginning of the simulation, the ratio is well below unity, and begins to increase, eventually becoming greater than unity at a time of around $t=25$. Finally, there are signs of a transition in the structure of the off-diagonal components of the electron pressure tensor. Initially in the force-free case, the structure and direction of the gradient in the vicinity of the X-point is more similar (qualitatively) to the constant guide field case, but at a later time in the evolution the structure looks more similar to the Harris case. It should be noted, however, that the transition we see is not as clear as going from purely guide field reconnection to purely anti-parallel reconnection, but instead we see initially primarily signatures of guide field reconnection, and signatures of both guide field and anti-parallel reconnection as the system evolves. This may be due to the fact that while $B_y$ at the dominant reconnection site decreases over time, it does not vanish completely, and there is a modified quadrupolar structure of $B_y$ at later times - not a transition to the quadrupolar structure seen in Harris sheet simulations. This could cause some features of guide field reconnection to persist, and is certainly a point open to further investigation. 

The dominant contribution to the reconnection electric field, $E_y$, was found to come from gradients of the off-diagonal components of the electron pressure tensor, in agreement wth previous findings for Harris and Harris plus constant guide field setups \cite{Hesse-1999,Kuznetsova-1998,Kuznetsova-2000,Kuznetsova-2001,Pritchett-2001,Hesse-2004,Ricci-2004,Pritchett-2005}. 


In this investigation, we have used only one set of parameters for the force-free run, which corresponds to a case where the ion distribution function is single-peaked in $v_y$, and has a double maximum in the $v_x$-direction, for small values of $z$ around zero. The electron distribution function is single-peaked in both $v_x$ and $v_y$. The distribution functions can of course both be single-peaked in $v_x$ for other sets of parameters, and can also have more pronounced double maxima in $v_x$, as well as a double maximum in $v_y$ \cite{Neukirch-2009}. A future study could investigate how the evolution of the system depends on the initial velocity space profile for this equilibrium. The dependence of the evolution on other parameters could be investigated, such as mass ratio, temperature ratio or initial current sheet thickness.


\acknowledgments{This project has received funding from the European Union's Seventh Framework Programme for research, technological development and demonstration under grant agreement SHOCK 284515 (FW \& TN). Website: project-shock.eu/home/. We also acknowledge financial support from the Leverhulme Trust, under grant no. F/00268/BB (CS, FW \& TN);  the U.K. Science and Technology Facilities Council via consolidated grant ST/K000950/1 (FW \& TN) and PhD studentship reference no. PPA/S/S/2005/04216 (MGH); and NASA's Magnetospheric Multiscale Mission (MH).}


\end{document}